\documentclass[prl, twocolumn, superscriptaddress]{revtex4-1}
\usepackage{graphicx}
\usepackage{amsmath,amssymb}
\usepackage{color}
\usepackage[colorlinks,
urlcolor=blue,
linkcolor=blue,
anchorcolor=blue,
citecolor=blue]{hyperref}
\renewcommand{\section}[1]{{\par\it #1.---}\ignorespaces}
\begin{document}
	\title{Non-Markovian effect on quantum optical metrology under dissipative environment}
	\author{Kai Bai}
	\affiliation{Key Laboratory of Artificial Micro- and Nano-structures of Ministry of Education and School of Physics and Technology, Wuhan University, Wuhan 430072, China}
	\author{Hong-Gang Luo}
	\affiliation{School of Physical Science and Technology \& Key Laboratory for Magnetism and Magnetic Materials of the MoE, Lanzhou University, Lanzhou 730000, China}
	\affiliation{Beijing Computational Science Research Center, Beijing 100084, China}
	\author{ Wenxian Zhang}
	\affiliation{Key Laboratory of Artificial Micro- and Nano-structures of Ministry of Education and School of Physics and Technology, Wuhan University, Wuhan, China}
	\author{Meng Xiao}
	\email{phmxiao@whu.edu.cn}
	\affiliation{Key Laboratory of Artificial Micro- and Nano-structures of Ministry of Education and School of Physics and Technology, Wuhan University, Wuhan, China}

	\begin{abstract}
		
	Quantum metrology utilizes quantum effects to reach higher precision measurements of physical quantities compared with their classical counterparts. However the ubiquitous decoherence obstructs its application. Recently, non-Markovian effects are shown to be effective in performing quantum optical metrology under locally dissipative environments [PhysRevLett.123.040402 (2019)]. However, the mechanism is still rather hazy.  Here, we uncover the reason  why forming a bound state can protect the quantumness against a dissipative ambient via the quantum Fisher information of entangled coherent states. An exact analytical expression of the quantum Fisher information in the long-encoding-time condition is derived, which reveals that the dynamics of precision can asymptotically reach the ideal-case-promised one easily when the average photon number is small. Meanwhile, the scaling exhibits a transition from the weak Heisenberg limit to the subclassical limit with increasing of average photon number. Our work provides a recipe to realize ultrasensitive measurements in the presence of noise by utilizing non-Markovian effects.	
		
	\end{abstract}

	\maketitle
	\section{Introduction}\label{introduction}
	Quantum metrology \cite{PhysRevLett.111.173601,Giovannetti1330,PhysRevLett.96.010401,RevModPhys.90.035005} is a very active research field that utilizes quantum effects, such as entanglement \cite{PhysRevLett.102.100401,Nagata726,PhysRevLett.112.103604,Luo620} and squeezing \cite{PhysRevD.23.1693,MA201189,PhysRevLett.118.140401}, to realize high-precision measurements of physical quantities. It is well-known that
	the enhanced precision due to the quantum effects can go far beyond their classical counterparts. In recent years, quantum metrology has been widely
	used in various directions such as gravitational wave detection \cite{PhysRevLett.110.181101}, radar \cite{PhysRevLett.114.080503}, magnetic field sensing \cite{Jones1166,PhysRevLett.112.150801,PhysRevLett.113.103004,11,RevModPhys.89.035002,Chalopin}, atomic clocks \cite{YeJ2014,PhysRevLett.117.143004,Hosten}. It also exhibits far-reaching impacts on quantum imaging \cite{1464-4266-4-3-372,PhysRevX.6.031033,PhysRevLett.117.190802}, optical lithography \cite{PhysRevLett.85.2733}, quantum information \cite{Quantuminformation},
	quantum biology \cite{doi:10.1063/1.4724105,PhysRevX.4.011017,Taylor20161}, etc.

	Physically,	a widely used quantum metrology scheme is based on the Mach-Zehnder interferometer (MZI). Here the benchmark compared against is the standard quantum limit (SQL) or  shot-noise limit (SNL), namely $N^{-1/2}$, where $N$ is the average photon number of the input state \cite{PhysRevLett.101.040403}. By utilizing the squeezed light, the precision was pointed out to surpass the SNL and eventually reach the Zeno limit (ZL)  $N^{-3/4}$ in 1981 by Caves \cite{PhysRevD.23.1693,HLL}. After that, various quantum states of light, such as the N00N state \cite{PhysRevLett.85.2733}, the twin Fock state \cite{PhysRevLett.71.1355}, the definite-photon-numbe state  \cite{PhysRevLett.102.040403,PhysRevA.80.013825}, the two-mode squeezed vacuum state \cite{PhysRevLett.104.103602} and the entangled coherent state (ECS) \cite{PhysRevLett.107.083601,PhysRevA.88.043832} have been extensively studied and exhibited exciting performances, which can reach or even surpass the weak Heisenberg limit (HL), i.e., $1/N$. Here we adopt the definition of weak Heisenberg limit from Ref. \cite{PhysRevA.95.032113}. However, these ideal-case-promised precision deteriorate quickly in experimental realization due to the unavoidable decoherence effects caused by the influence of environments \cite{PhysRevA.81.033819,PhysRevA.90.033846,PhysRevLett.102.040403,PhysRevA.81.033819,PhysRevA.80.013825,PhysRevA.95.053837,PhysRevA.78.063828,Gilbert:08,PhysRevA.75.053805,ZhangWP,ysw}. Previously, the decoherence of the optical probe was phenomenologically desccribed by a transmissivity \cite{PhysRevLett.102.040403,PhysRevLett.107.083601,PhysRevA.88.043832,PhysRevLett.102.040403,PhysRevA.80.013825,NaturePhysics7406-411,NatureCommunications31063}. That is to say, we are considering a continuous photon loss model using a Born-Markovian master equation \cite{PhysRevLett.108.130402,Lu2015}. Under such a description, the precision eventually gets worse and worse for long-encoding-time running. However, such a treatment was later found insufficient in real physical systems \cite{Guo2011,Bernardes2015,Groblacher2015,Liu2016,Krinner2018,PhysRevA.85.060101,PhysRevLett.108.210402,PhysRevLett.109.170402,PhysRevA.86.010102,PhysRevLett.107.080404,RevModPhys.88.021002,Rivas_2014,LI20181,ysw,PhysRevLett.109.233601,PhysRevLett.123.040402,PhysRevLett.116.120801,PhysRevA.92.010102}.  
	 Intriguingly, with the squeezed states as input and the photon difference measurement, the ZL can even be asymptotically recovered by utilizing the non-Markovian effect and with the aid of a bound state \cite{PhysRevLett.123.040402}. That work first pointed out the fact that non-Markovian effects can be effective in performing quantum optical metrology. However, 
	the detailed contribution of the non-Markovian effect and the mechanism behind remain blurred because there the measurement scheme cannot exploit all the available information of sensitivity.  As an example, it is not clear whether the transition from the ZL to the SNL at large $N$ is due to the specific measurement scheme or the non-Markovian effects. More importantly, although the ZL has been reached in Ref. \cite{PhysRevLett.123.040402}, whether the fascinating HL can be surpassed remains unknown.

	\begin{figure}[tbp]
		\centering
		\includegraphics[width=3.6cm,height=5.85cm]{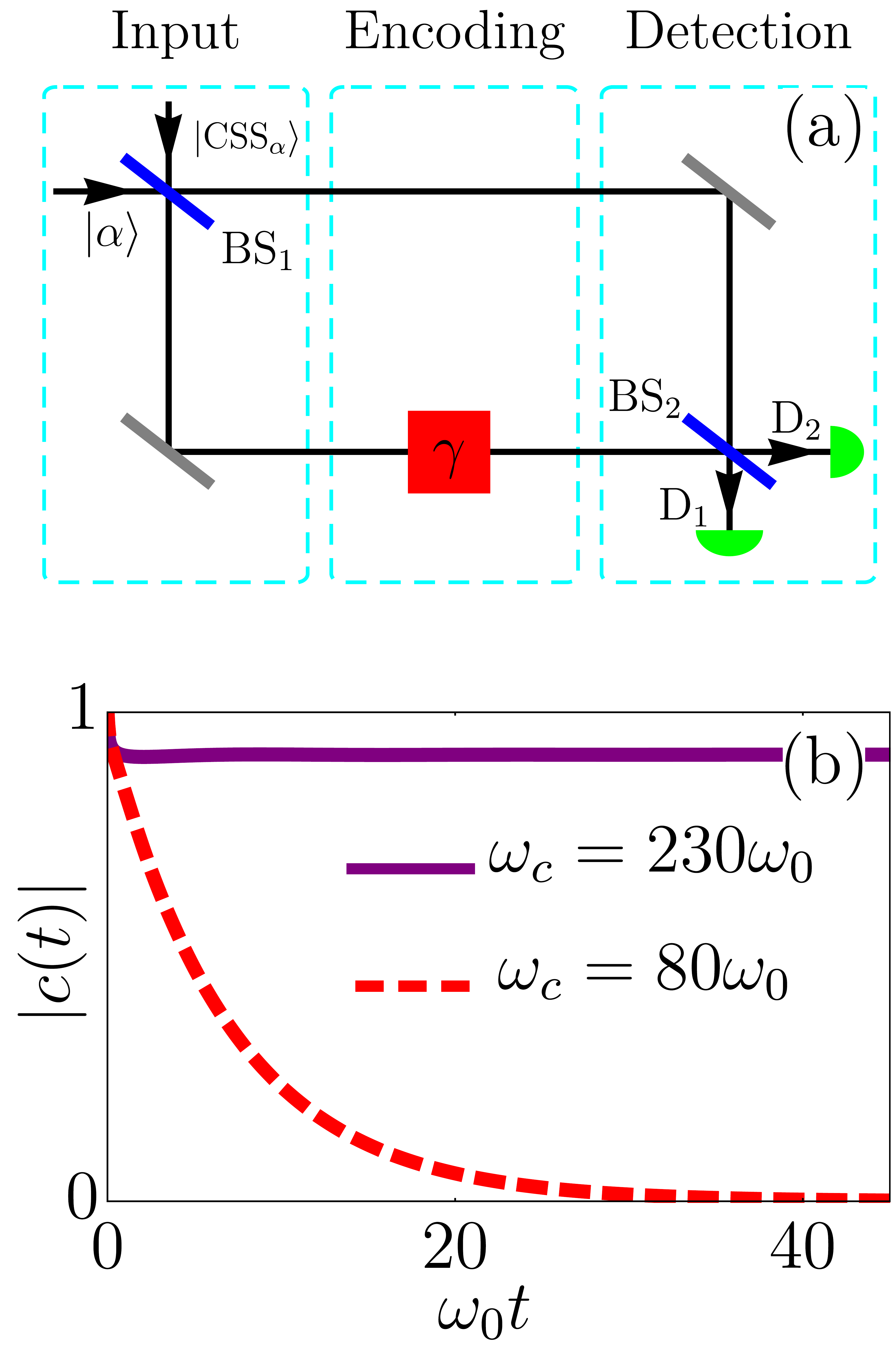}~~\includegraphics[width=4cm,height=5.6cm]{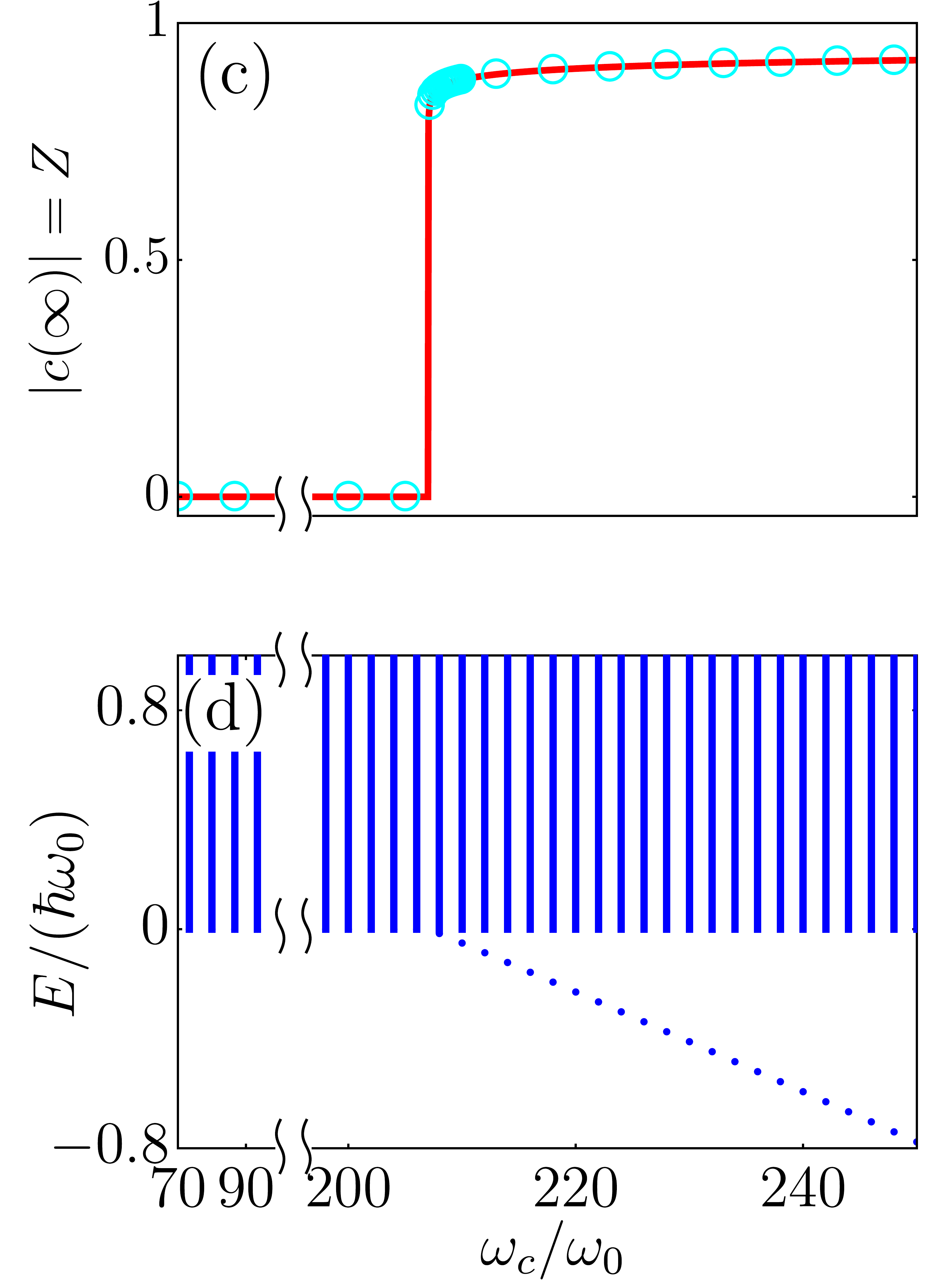}\\
		\caption {(a) Schematic diagram of the MZI-based quantum metrology. Two input fields ($|\text{CSS}_{\alpha}\rangle$ and $|\alpha\rangle$) interact at a beamsplitter $\text{BS}_1$, become ECSs and propagate along two arms. One of the fields couples with a system where the estimated parameter $\gamma$ is encoded. After interfering at $\text{BS}_2$, the output fields are measured to finish a typical parameter estimation. (b) The evolution of $|c(t)|$ with two different $\omega_{c}$s. The red dashed (purple) line corresponds to the case where a bound state is absent (present). (c) Long-time behavior of $|c(t)|$ (cyan circles) as a function of $\omega_{c}$. The results agree well with $Z$ from the bound-state analysis  (red solid line). (d) Energy spectra of the whole system including the optical field and its environment. For panels (b)-(d), we set $s=1$, $\gamma=\pi \omega_{0}$ and $\eta=0.02$.  }
		\label{MZ}
	\end{figure}

     In this paper, we investigate the contribution of non-Markovian effects on optical quantum metrology by calculating  the quantum Fisher information (QFI) which can perfectly characterize the sensitivity of a state with respect to the change of a specific parameter \cite{PhysRevLett.72.3439,BRAUNSTEIN1996135}. For the input state, we choose the ECS, one of the quantum states which can beat the weak HL under ideal conditions. We find that the dynamics and scaling of precision are closely related to the frequency of photons and the estimated parameter are dramatically affected by the non-Markovian effect. With the aid of a bound state, an exact analytical expression of quantum Fisher information in the long-encoding-time condition is derived. It reveals that the precision can surpass the weak HL and asymptotically approach the ideal-case-promised one with the increase of $Z$ [see Eq. \eqref{defZ} for the definition of $Z$]. Meanwhile, the scaling exhibits a transition from the weak HL to the sub-SNL. Our work points out that non-Markovian effects with the aid of a bound state can suppress dissipation very well, and its ability to protect quantumness does not change with the encoding time. Hence the precision get better and better with the increase of time. However, this ability to protect quantumness weakens quickly with the increase of $N$.

	\section{Ideal quantum metrology}
	Generally, the process of quantum metrology based on the MZI consists of three steps as shown  schematically in Fig. \ref{MZ}(a).   Firstly, the input states $|\Psi_{\text{in}}\rangle$ are prepared in two different modes labeled by 1 and 2.  After interaction at the first $50-50$ beam splitter ($\text{BS}_{1}$) $\hat{V}_{1}$, the beam along one arm interacts with the system to encode the parameter information while the beam along the other arm remains intact. The encoding process is governed by the evolution operator $\hat{U}_{0}(\gamma,t)=\text{exp}(-i\hat{H}_{0}t/\hbar)$ with
	\begin{equation}
	\hat{H}_0=\hbar\omega_0\sum_{m=1,2}\hat{a}_m^\dag\hat{a}_m+\hbar\gamma{(\hat{a}_2^\dag\hat{a}_2)}^k.
	\end{equation}
	Here $\gamma$ is the parameter under estimation,  $k$ is the order parameter of nonlinearity (here we set $k=1$, implying a conventional phase difference $\gamma t$) \cite{PhysRevLett.107.083601,PhysRevA.86.043828}, and $\hat{a}_m^\dag$ ($\hat{a}_m$) denotes the creation (annihilation)  operator of mode $m$ \cite{PhysRevLett.107.083601,PhysRevLett.101.040403,PhysRevA.77.012317}. In the detection part, beams along two different arms interact at the second $50-50$ beam splitter ($\text{BS}_2$) $\hat{V}_{2}$, and measurements are performed thereafter.  
	
	The highest possible precision of the estimated parameter is given by the quantum Cram\'{e}r-Rao bound \cite{PhysRevLett.100.073601,PhysRevLett.72.3439,BRAUNSTEIN1996135} of the state  $|\Psi_{\text{end}}\rangle\equiv\hat{U}_{0}(\gamma,t)\hat{V}_{1}|\Psi_{\text{in}}\rangle$. Here the subscript end denotes
	the end of the parameter encoding process. The quantum Cram\'{e}r-Rao bound is 
	\begin{equation}\label{CRB}
	\delta\gamma\geq\frac{1}{\sqrt{\mu F_{Q}}},
	\end{equation}
	where $\mu=1$ here is the number of experimental runs\cite{PhysRevLett.105.180402}, and $F_Q$ denotes the QFI \cite{PhysRevA.81.043624,PhysRevLett.102.040403,PhysRevA.80.013825,PhysRevA.88.043832}. For a pure state, the QFI is given by
	\begin{equation}\label{PQF}
	F_{Q}=4[\langle\Psi_{\text{end}}^{\prime}|\Psi_{\text{end}}^{\prime}\rangle-|\langle\Psi_{\text{end}}^{\prime}|\Psi_{\text{end}}\rangle|^{2}],
	\end{equation} where $|\Psi_{\text{end}}^{\prime}\rangle=\partial|\Psi_{\text{end}}\rangle/\partial\gamma$. For a mixed state described by a density matrix $\rho_{\text{end}}$, $F_Q$ is given by 
	\begin{equation}\label{mixedf}
	F_{Q}=\sum_{i,j}\frac{2}{\lambda_{i}+\lambda_{j}}|\langle\lambda_{i}|\rho_{\text{end}}^{\prime}|\lambda_{j}\rangle|^{2},
	\end{equation}
	where $\rho_{\text{end}}^{\prime}=\partial\rho_{\text{end}}/\partial\gamma$, $|\lambda_{i}\rangle$ denotes the orthonormalized eigenvector of $\rho_{\text{end}}$ with the corresponding eigenvalues $\lambda_{i}$, and the summation is over all the eigenstates.

We consider the case where the input states are a coherent state $|\alpha\rangle$  and a coherent state superposition given by $|\text{CSS}_{\alpha}\rangle=N_{\alpha}(|\alpha\rangle+|-\alpha\rangle)$ with $N_{\alpha}=\text{[}2(1+e^{-2|\alpha|^{2}})]^{-1/2}$ \cite{PhysRevA.86.043828}. After interacting at the first beamsplitter $\hat{V}_{1}=\text{exp}[\frac{\pi}{4}(\hat{a}_{1}^{\dagger}\hat{a}_{2}-\hat{a}_{2}^{\dagger}\hat{a}_{1})]$, the input states become the ECS given by
\begin{equation}
|\text{ECS}\rangle=\mathcal{N}_{\alpha}[|\alpha,0\rangle+|0,\alpha\rangle],
\end{equation}
where $|\alpha,0\rangle\equiv|\alpha\rangle_{1}|0\rangle_{2}$  with $|0\rangle_i$ and $|\alpha\rangle_i$ describing respectively, the vacuum and coherent states in spatial mode $i$, and $\mathcal{N}_{\alpha}=[2(1+e^{-|\alpha|^{2}})]^{-1/2} $ is the normalization constant \cite{PhysRevLett.107.083601}.
 The total average photon number of the input states here is given by $N=|\alpha|^{2}/(1+e^{-|\alpha|^{2}})$. Since here we are working with pure states, combining Eqs. \eqref{CRB} and \eqref{PQF}, we have   
\begin{eqnarray}\label{idealF}
F_{Q}&&=2Nt^{2}[1+\mathcal{W}(Ne^{-N})]+N^{2}t^{2},\\
\delta\gamma&&\geq[2Nt^{2}[1+\mathcal{W}(Ne^{-N})]+N^{2}t^{2}]^{-1/2},
\end{eqnarray}
where $\mathcal{W}(x)$ denotes the Lambert $W$ function. It is easy to see that the measurement precision beats both the SNL and weak HL, manifesting the significance of the entanglement of light in the metrology scheme.

\section{Effects of dissipative noises}
In practice, the decoherence caused by the unavoidable interaction of the probe with the environment is a major obstacle for quantum metrology. Conventionally, the decoherence of the optical probe based on the MZI is analyzed in a phenomenological manner by introducing a transmissivity \cite{PhysRevA.81.033819,PhysRevA.90.033846,PhysRevLett.102.040403,PhysRevA.81.033819,PhysRevA.80.013825,PhysRevA.95.053837,PhysRevA.78.063828,Gilbert:08,PhysRevA.75.053805,ZhangWP}. That is to say, we are considering a continuous photon loss model using a Born-Markovian master equation
\cite{PhysRevLett.102.040403,PhysRevLett.108.130402,Lu2015}. However, with the rapid development of quantum reservoir engineering technology, experimentalists get access to engineering the environmental spectral densities such that the non-Markovian effects during the decoherence process can be observed. The interplay between the system and its environment caused by inherent non-Markovian nature induces diverse new characters as have been explored in various systems \cite{Guo2011,Bernardes2015,Groblacher2015,Liu2016,Krinner2018,PhysRevA.85.060101,PhysRevLett.108.210402,PhysRevLett.109.170402,PhysRevA.86.010102,PhysRevLett.107.080404,RevModPhys.88.021002,Rivas_2014,LI20181}. 
Here we focus on the non-Markovian effects on quantum optical metrology under the long-encoding-time condition.

With the dissipative noise taken into consideration, the Hamiltonian governing the parameter encoding now reads
\begin{equation}
\hat{H}=\hat{H}_0+\hbar\sum_k[\omega_k\hat{b}_k^\dag\hat{b}_k+g_k(\hat{a}_2\hat{b}_k^\dag+\text{H.c.})],
\end{equation}
where $\hat{b}_{k}^{\dagger}$ is the creation operator of the $k$th environmental mode with frequency $\omega_{k}$, and $g_{k}$ is the probe-environment coupling strength. Note here that the noise is introduced only in the process of encoding. This coupling can be characterized by the spectral density function $J(\omega)\equiv\sum_{k}|g_{k}|^{2}\delta(\omega-\omega_{k})$, which can be properly engineered \cite{Guo2011,Bernardes2015,Groblacher2015,RevModPhys.88.021002,Rivas_2014,LI20181}. In the continuum limit, it reads $J(\omega)=\eta\omega(\frac{\omega}{\omega_{c}})^{s-1}e^{-\frac{\omega}{\omega_{c}}}$, where $\eta$ and $\omega_{c}$ represent the coupling constant and cutoff frequency, respectively. The noise is classified as sub-Ohmic with $0<s<1$, Ohmic  with $s=1$, and super-Ohmic  with $s>1$ \cite{RevModPhys.59.1}.

Assuming that the environment is initially in the vacuum state $|\Psi_\text{E}(0)\rangle=|\{0_k\}\rangle$, and the exact master equation governing the process of parameter encoding can be derived by the Feynman-Vernon's influence functional method \cite{PhysRevE.90.022122,FEYNMAN1963118,PhysRevA.76.042127,PhysRevA.82.012105} as
\begin{eqnarray}
 \dot{\rho}(t)	=&&-i\omega_{0}[\hat{a}_{1}^{\dagger}\hat{a}_{1},\rho(t)]-i\Omega(t)[\hat{a}_{2}^{\dagger}\hat{a}_{2},\rho(t)]\nonumber\\
 &&+\Gamma(t)[2\hat{a}_{2}\rho(t)\hat{a}_{2}^{\dagger}-[\rho(t),\hat{a}_{2}^{\dagger}\hat{a}_{2}]_{+}],
\end{eqnarray}
where $[\rho(t),\hat{a}_{2}^{\dagger}\hat{a}_{2}]_{+}\equiv\rho(t)\hat{a}_{2}^{\dagger}\hat{a}_{2}+\hat{a}_{2}^{\dagger}\hat{a}_{2}\rho(t)$, the renormalized frequency $\Gamma(t)$ and the decay rate $\Omega(t)$ satisfy $\Gamma(t)+i\Omega(t)=-\dot{c}(t)/c(t)$ with $c(t)$ determined by
\begin{eqnarray}\label{c(t)}
\dot{c}(t)+i(\gamma+\omega_{0})c(t)+\int_{0}^{t}f(t-\tau)c(\tau)d\tau=0
\end{eqnarray}
and under the initial condition $c(0)=1$. Here $f(t-\tau)\equiv\int J(\omega)e^{-i\omega(t-\tau)}d\omega$
is the environmental correlation function. Equation \eqref{c(t)} incorporates all the non-Markovian effects between the probe and environment into self-consistent coefficients $\Omega(t)$ and $\Gamma(t)$. For any quantum states, the transmission rate of photons is $N(t)/N=[1+|c(t)|^{2}]/2$.

For the ECSs under consideration, 
\begin{eqnarray}\label{rho(t)}
\rho(t)=&&\mathcal{N}_{\alpha}^{2}\{|\alpha e^{-i\omega_{0}t},0\rangle\langle\alpha e^{-i\omega_{0}t},0|\nonumber\\&&+e^{-\frac{|\alpha|^{2}}{2}(1-|c(t)|^{2})}[|\alpha e^{-i\omega_{0}t},0\rangle\langle0,c(t)\alpha|+\text{H.c.}]\nonumber\\&&+|0,c(t)\alpha\rangle\langle0,c(t)\alpha|\},
\end{eqnarray} wherein the second term (the "coherence" term) denotes the quantum effect with a factor of $\text{exp}[-\frac{|\alpha|^{2}}{2}(1-|c(t)|^{2})]$. Combined with Eq. \eqref{mixedf}, the QFI can thus be
evaluated from Eq. \eqref{rho(t)}. The end result is analytically complicated and suitable only for numerical evaluation. However, the asymptotic behavior of the QFI in the long-encoding-time limit can be obtained analytically through analyzing Eq. \eqref{c(t)}.

 When the coupling between the probe and the environment is weak and the typical time scale of $f(t-\tau)$ is much smaller than that of the input field, the Markovian approximation can be applied in Eq. \eqref{c(t)}. Then one can have $c(t)=\exp\left\{ -\left[\kappa+i\left(\omega_{0}+\gamma+\Delta\right)\right]t\right\} $ with $\kappa=\pi J(\omega_{0}+\gamma)$ and $\Delta=\mathcal{P}\int_{0}^{\infty} J(\omega)/(\omega-\omega_{0}-\gamma)d\omega$ \cite{PhysRevE.90.022122}, where $\mathcal{P}$ represents the principal integration. Thus with Eqs. \eqref{rho(t)} and \eqref{mixedf}, $F_{Q}\approx2Nt^{2}e^{-2\kappa t}$.
The best measuring precision defined in Eq. \eqref{CRB} is $\text{min}(\delta\gamma)\thickapprox e\kappa(2N)^{-1/2}$ at $t\thickapprox\kappa^{-1}$. This conclusion is consistent with Ref. \cite{PhysRevLett.123.040402}. 

In the non-Markovian dynamics, Eq. \eqref{c(t)} can be linearized by a Laplace transform $\tilde{c}(p)=[p+i(\omega_{0}+\gamma)+\int_{0}^{\infty} J(\omega)/(p+i\omega)d\omega]^{-1}$. Then $c(t)$ is obtainable by inverse Laplace transform of $\tilde{c}(p)$ via finding its poles from\begin{equation}\label{root}
y(\varpi)\equiv\omega_{0}+\gamma-\int_{0}^{\infty}\frac{J(\omega)}{\omega-\varpi}d\omega=\varpi,~(\varpi=ip).
\end{equation}It is interesting to note that  $\hbar\varpi$ is the eigenenergy of the local system in the single-excitation subspace. To uncover this point, we first expand the eigenstate as $|\Phi\rangle=(x\hat{a}^\dag_2+\sum_ky_k\hat{b}^\dag_k)|0_2,\{0_k\}\rangle$. From the stationary Schr\"{o}dinger equation of the local system, we have $[E-\hbar(\omega_0+\gamma)]x=\sum_k\hbar g_ky_k$ and $y_k=\hbar g_k x/(E-\hbar\omega_k)$ with $E$ being its eigenenergy. This immediately leads to Eq. \eqref{root} with the replacement of $\varpi$ by $E/\hbar$. A direct indication from Eq. \eqref{root} is that, the decoherent dynamics of the probe is essentially determined by the single-excitation subspace. Noted that $y(\varpi)$ is a monotonically decreasing function of $\varpi$ in the regime $\varpi<0$, Eq. \eqref{root} possesses one and only one isolated root $\varpi_\text{b}$ in the regime $\varpi<0$ provided $y(0)<0$. Meanwhile, Eq. \eqref{root} has infinite roots in the regime $\varpi>0$, which form a continuous energy spectrum. Each root of Eq. \eqref{root} corresponds to an eigenstate of the system. Hence the isolated root corresponds to a bound state because its eigenenergy $\hbar\varpi_\text{b}$ falls out of the energy band (continuous energy spectrum) \cite{PhysRevA.81.052330}. 

The formation of this bound state has profound influences on the decoherent dynamics of the probe. This can be visualized by making the inverse Laplace transform
\begin{eqnarray}
c(t)=Ze^{-i\varpi_\text{b}t}+\int_{i\epsilon+0}^{ i\epsilon+\infty}{d\varpi\over 2\pi}\tilde{c}(-i\varpi)e^{-i\varpi t},\label{invlpc}
\end{eqnarray}
where
\begin{eqnarray}
 Z=[1+\int_{0}^\infty{J(\omega)/(\varpi_\text{b}-\omega)^2}d\omega]^{-1}  \label{defZ}
\end{eqnarray}  
 and the second term represents the contribution of the continuous spectrum.  Oscillating in time with continuously changing frequencies, the second term decays to zero in the long-time limit due to the out-of-phase interference. Therefore, if there is no bound state, then $\lim_{t\rightarrow \infty}c(t)=0$ indicates a complete decoherence. On the other hand, if the bound state is formed, 
 then $\lim_{t\rightarrow \infty}c(t)=Ze^{-i\varpi_\text{b}t}$ implies a decoherence-suppression.  For the Ohmic-type spectral density, it can be proved that, a bound state is formed provided that $\omega_0+\gamma-\eta\omega_c \text{G}(s)\leq0$, where $\text{G}(s)$ is the Euler Gamma function.
 And for Ohmic spectral, the dependence of $Z$ on those parameters are given by $Z=[(\omega_{0}+\gamma-\eta\omega_{c})/\varpi_{\text{b}}+(\varpi_{\text{b}}-\omega_{0}-\gamma)/\omega_{c}]^{-1}$.
  Bound states also play dominate role in noncanonical thermalization \cite{PhysRevE.90.022122} and quantum-correlation preservation \cite{PhysRevA.88.012129}.

 We focus on the case where a bound state is present. Substituting the asymptotic solution $c(t)=Ze^{-i\varpi_\text{b}t}$ into Eq. \eqref{mixedf}, we obtain
\begin{eqnarray}\label{QFI}
F_{Q}=&&2t^{2}Z^{4}N+2t^{2}Z^{6}N\mathcal{W}(Ne^{-N})\nonumber\\&&+e^{-|\alpha|^{2}(1-Z^{2})}t^{2}Z^{6}N^{2},
\end{eqnarray}where the dependence of $\varpi_\text{b}$ on $\gamma$ has been taken care via $\partial_{\gamma}\varpi_{\text{b}}=Z$.  Equation \eqref{QFI} gives the central result of this paper. When $Z=1$, it reduces to Eq. \eqref{idealF} in the ideal case. For a general case, $Z$ depends on both $\omega_{0}$ and $\gamma$ according to Eqs. \eqref{root} and \eqref{defZ}. This reveals that one can tune the working frequency of the quantum probe to greatly improve the precision, this information can not draw from from  Refs. \cite{PhysRevLett.107.083601,PhysRevA.88.043832}. Meanwhile, due to the presence of a bound state, the benchmark for precision decreases with the increase of encoding time, which is in sharp contrast to previous works \cite{PhysRevA.81.033819,PhysRevA.90.033846,PhysRevLett.102.040403,PhysRevA.80.013825,PhysRevA.95.053837,PhysRevA.78.063828,PhysRevA.75.053805,PhysRevLett.108.130402,Lu2015,ZhangWP} and can be seen in Fig. 2(a). Even more intriguingly, in the long-time limit, $\text{min}(\delta\gamma)$ asymptotically approaches the ideal-case-promised precision when $Z$ approaches 1. In other words, when $N$ is small, the weak HL can be achieved and even surpassed easily under a real noisy environment. Moreover, we also notice that the scaling with $N$ exhibits a transition from the weak HL to the sub-SNL. This can be seen from the fact that the first two terms in Eq. \eqref{QFI} are proportional to $N$ and dominant for large enough $N$ as $|\alpha|^{2}\approx N$.   
 
The above discussed phenomena are attributed to the formation of a bound state and its ability to protect quantumness. In detail, the factor $e^{-\frac{1}{2}(1-|Z|^{2})|\alpha|^{2}}$in Eq. \eqref{rho(t)} [$\lim_{t\rightarrow \infty}|c(t)|=Z$] characterizes the ability of the bound state in protecting quantumness. Such a factor does not change with $t$, but decreases sharply with the increase of $N$, which can explain why there is a transition from the weak HL to the sub-SNL. Meanwhile, we point out that the transmission rate of photons $N(t)/N=(1+|Z|^{2})/2$, does not change with $t$ and $N$ in the presence of a bound state and the long time limit. This fact further proves that the previous Markovian approximation treatment is insufficient.

\begin{figure}
	\centering
	\includegraphics[width=1\columnwidth]{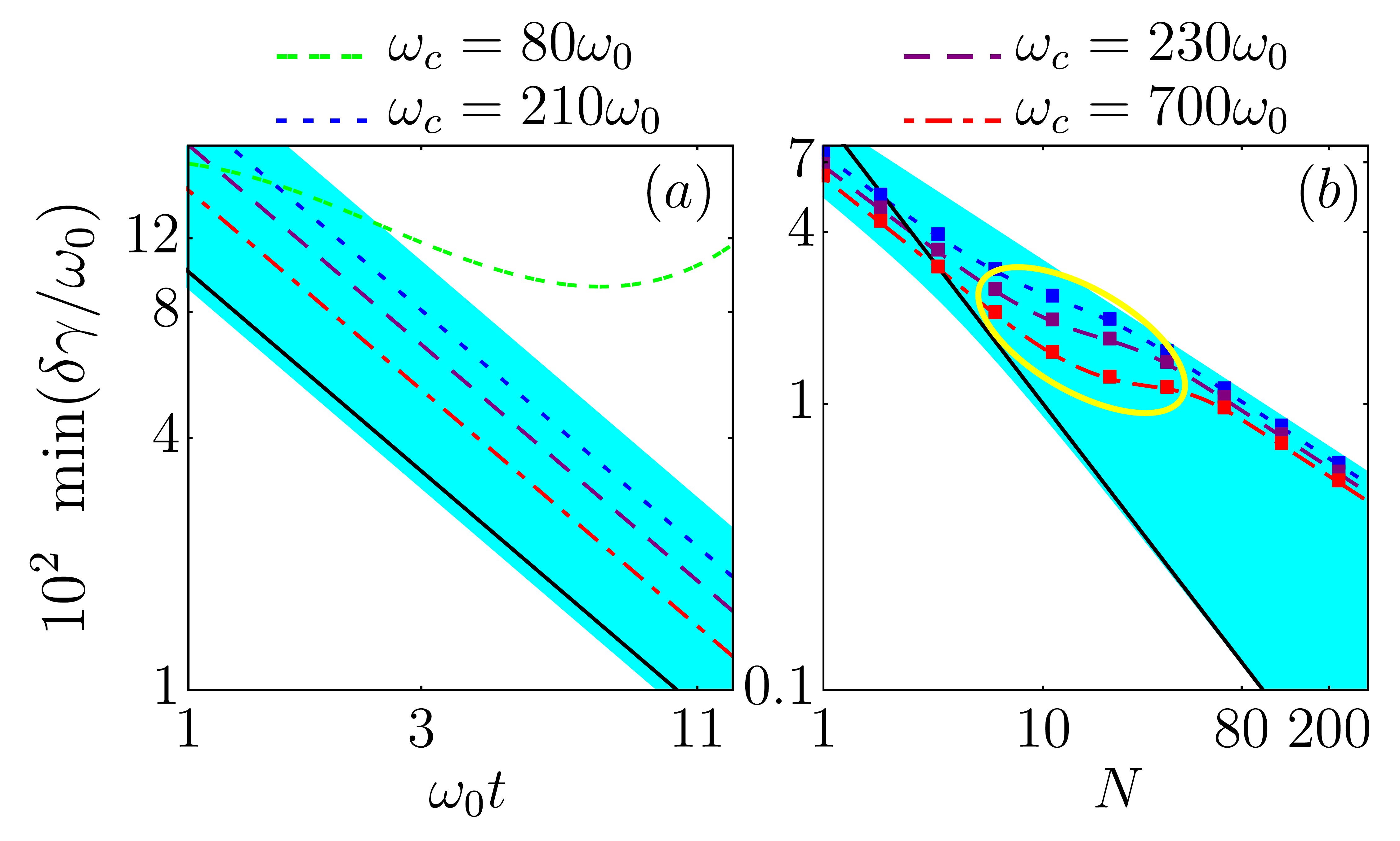}\\
	\caption { $\text{min}(\delta\gamma)$ as a function of (a) time  and (b) average photon number with different $\omega_c$s. The black lines denote the weak HL. The shaded areas in panels (a) and (b) are bound by the SNL (upper bound) and the ideal-case-promised precision (lower bound) in the absence of noise. $N=10$ in panel (a), $t=10\omega_0^{-1}$ in panel (b), and the other parameters used are the same as those in Fig.\ref{MZ}.}
	\label{rt}
\end{figure}

\section{Numerical results} We now proceed to numerically solve the dynamical equations to supplement our analysis above.  
 
\begin{figure}
	\centering
	\includegraphics[width=1.0\columnwidth]{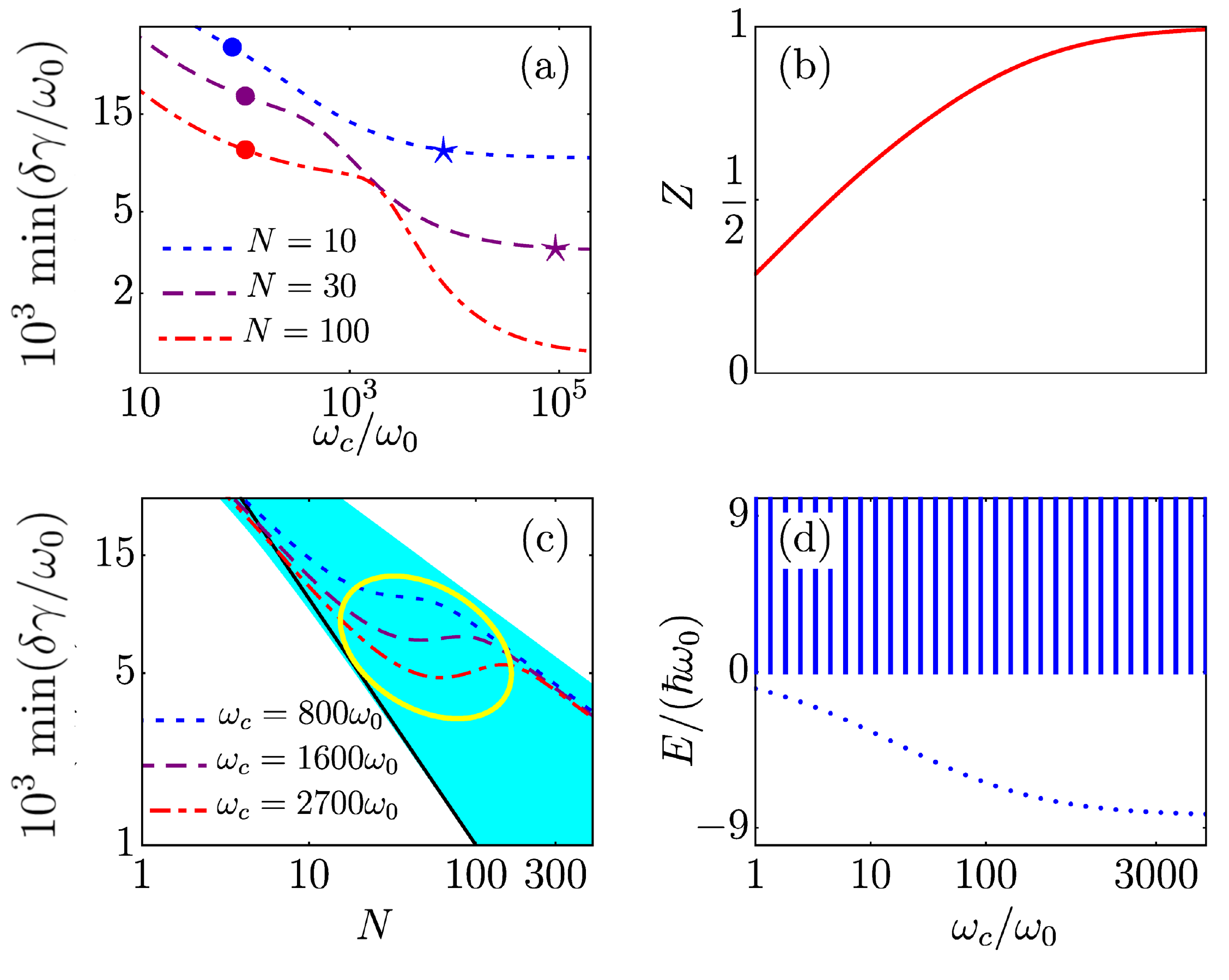}\\
	\caption {
		(a) $\text{min}(\delta\gamma)$s as functions of $\omega_{c}$  for different average photon numbers, where the circles and asterisks denote the corresponding SNLs and weak HLs. (b) $Z$ as a function of $\omega_{c}$. (c) $\text{min}(\delta\gamma)$s as functions of the average photon number $N$ for different $\omega_{c}$s, where the shaded area in panel (c) is bound by the SNL (upper bound) and the ideal-case-promised precision (lower bound) in the absence of noise, and the yellow ellipse highlights the transition region. The black line denotes the weak HL. (d) Energy spectra of the whole system including the optical field and its environment. Here we set $\eta=3(\omega_{0}+\gamma)/[\omega_{c}G(s)]$, $t=10\omega_{c}^{-1}$, and the other parameters are the same as those in Fig. \ref{MZ}.
	}
	\label{fig3}
\end{figure}

To verify the distinguished role played by the bound state, in Fig.  \ref{MZ}(b) we plot the evolution of $|c(t)|$ with different $\omega_{c}$s, which exhibit two distinct dynamics. These can be seen more clearly in Fig.\ref{MZ}(c), where the long-time behavior $|c(\infty)|$ (the open cyan disks) shows an abrupt jump from zero to a finite value with the increase of $\omega_c$. This jump matches exactly with the formation of a bound state as can be seen from the eigenspectra in Fig. \ref{MZ}(d). This interesting feature actually originates from the fact that a bound state, as a stationary state of the local system, preserves quantum coherence during time evolution. We also numerically show that $|c(\infty)|$ exactly coincides with the bound-state analysis $Z$ (the red line). Later we see that $|c(\infty)|=Z$ approaching 1 can be approximately achieved with the increase of $\omega_c$. In such a case, the precision returns to the ideal case.

 With  Eq. \eqref{c(t)} solved, the exact precision $\text{min}(\delta\gamma)$ can be obtained by using Eqs. \eqref{CRB}, \eqref{mixedf} and \eqref{rho(t)}. The dynamics of $\text{min}(\delta\gamma)$ is shown in Fig. \ref{rt}(a). 
 It is clear that, $\text{min}(\delta\gamma)$ increases with time after a decrease without the formation of a bound state, which is consistent with the Markovian case (the noisy metrology scheme performs worse and worse with  increasing of encoding time). However, as long as a bound state is formed, $\text{min}(\delta\gamma)$ becomes a monotonically decreasing function of $t$. Figure \ref{rt}(a) also reveals that, not only can the SNL be surpassed, but also the ideal-case-promised precision can even be asymptotically retrieved with the increase of $\omega_{c}$. However, it has to be noticed that the ideal-case-promised precision is difficult to retrieve when $N$ is large. To show this point, we plot $\text{min}(\delta\gamma)$ as a function of $N$ in Fig. \ref{rt}(b). It is clear that $\text{min}(\delta\gamma)$ exhibits a transition from the weak HL to the sub-SNL as highlighted by the yellow ellipse. Note here that the transition from the weak HL to the SNL depends on the value of $\omega_{c}$. This transition is consistent with the analysis of Eq. \eqref{QFI}, of which the results are also given by the solid squares in Fig. \ref{rt}(b).

 To further illustrate the point that the weak HL can be approached and even surpassed, we show $\text{min}(\delta\gamma)$ as a function of $\omega_{c}$ in Fig. \ref{fig3}(a). Meanwhile, Fig. \ref{fig3}(b) clearly shows, with the aid of a bound state [see Fig. \ref{fig3}(d) for eigenspectra], $Z$ can approach 1 with the increase of $\omega_{c}$. They validate again the statements analyzed before in Eq. \eqref{QFI}. Moreover, 
 there is an interesting minimum in the transition region of scaling when $\omega_{c}>1092\omega_{0}$ or $Z>0.9625$ [see Fig. \ref{fig3}(c)] inside the yellow ellipse, which contrasts sharply with the monotonically decreasing performance. This phenomenon  will be further studied in the future works.

\section{Conclusions and Discussions}\label{conclusion} 
In summary, we have studied the non-Markovian noise effect on quantum optical metrology by exactly solving the QFI of ECSs in a locally dissipative environment. Under the long-encoding-time condition, an exact analytical expression of the QFI is derived to show the role of non-Markovian effects on the ultimate phase sensitivity.  We reveal that, in the presence of noise, the non-Markovian effects with the aid of a bound state make the superiority of $t$ as a resource recovered; However, the contribution of $N$ weakens quickly with the increase of $N$. On one hand, our results illustrate the non-Markovian effect on quantum optical metrology, enriching the understanding of the ECSs. On the other hand, our work supplies a guideline to realize the ultrasensitive measurements in practice  through forming a bound state by engineering the reservoir.

Note that, although only the class of Ohmic spectral density is considered in this work, our results can be generalized to other types of spectral densities. Although the bound state formation condition and the relation between $Z$ and other parameters are different from spectral density to spectral density,  Eq. (15) preserves the same, and thus we can always exploit the non-Markovianity by solving the Eq. (13) to determine how to adjust a certain parameter. With the rapid development of the reservoir engineering technique, many structured environments with controllable spectral densities have been realized \cite{ER1,Kienzler53}. Based on the MZI, non-Markovian effects have been observed in linear optical systems \cite{Guo2011,Bernardes2015}. An all-optical non-Markovian quantum simulator has been proposed \cite{PhysRevA.91.012122}. A sub-Ohmic spectrum has been engineered and the non-Markovian effect is observed in a micromechanical system \cite{Groblacher2015}, which as a quantum continuous-variable bosonic system shares similar characteristics with the quantum optical probe in the  quantum metrology setup studied in our system. The bound state has already been observed in a photonic- band-gapped environment in a circuit QED system \cite{Liu2016}. The Ohmic spectral density is possible to be controlled in a trapped ion system \cite{PhysRevA.78.010101}. All these systems supply potential experimental platforms to verify the phenomena discussed in our work.

We need to point out that the value of $\omega_{c}$ becomes impractically large for a large $N$,  as shown in Fig. \ref{fig3}(a). This is actually not a fatal problem for our scheme. The core objective of quantum metrology is to minimize $\delta\gamma$ with a given resource: the time  and the total average photon number. With fixed time $t$ and total average photon number, one can always increase $\mu$ in Eq. \eqref{CRB} to reduce $N$ of the input state (the total photon number is given by $\mu N$) so that the single experimental run is in the sub-HL region. Meanwhile, one can also engineer the environment structure to release the stringent requirement on $\omega_{c}$. For example, instead of Ohmic spectral density, one can consider super Ohmic spectral density.

\section{Acknowledgments}
The authors thank Zhen Peng and Can Shao for helpful discussions. The numerical calculations in this paper have been done on the supercomputing system in the Supercomputing Center of Wuhan University. The work is supported by the startup funding of Wuhan University and by the National Natural Science Foundation (Grants No. 11904264, No. 11674139, No. 11834005, No.91836101,and No.U1930201).


\begin{thebibliography}{86}
\bibitem [{HLL()}]{HLL}%
\BibitemOpen
\href@noop {} {\ \bibinfo {series} {When the average photon number of the
		squeezed state equals that of the coherent state, it can be proved that the
		optimal limit with the squeezed state is the weak HL provided that a full
		knowledge of the photon number fluctuation is known
		\cite{PhysRevLett.100.073601}.}}\BibitemShut {Stop}%
\end{thebibliography}
%

\end{document}